\documentclass[pra,superscriptaddress,showpacs,twocolumn,amsmath,amssymb]{revtex4}% Physical Review A:twocolumn

\usepackage{graphicx}% Include figure files
\usepackage{bm}% bold math
\usepackage{epsfig}

\newcommand {\Fig}[1] {Figure~\ref{#1}}
\newcommand{\beq}{\begin{equation}}
\newcommand{\eeq}{\end{equation}}

\newcommand{\natc}{N@C$_{60}$}
\newcommand{\natcseventy}{N@C$_{70}$}

\newcommand{\csixty}{C$_{60}$}

\newcommand{\cstwo}{CS$_2$}
\newcommand{\stwocltwo}{S$_2$Cl$_2$}

\newcommand{\beqa}{\begin{eqnarray}}
\newcommand{\eeqa}{\end{eqnarray}}
\newcommand{\CPL}{Chem. Phys. Lett.}
\newcommand{\IJTP}{Int. J. Thermophys.}
\newcommand{\JCP}{J. Chem. Phys.}

\newcommand{\JACS}{J. Am. Chem. Soc.}
\newcommand{\JMOLLIQ}{J. Mol. Liq.}

\newcommand{\JPC}{J. Phys. Chem.}
\newcommand{\JPCM}{J. Phys. Cond. Matt.}

\newcommand{\PR}{Phys. Rev.}
\newcommand{\PRA}{Phys. Rev. A}

\newcommand{\ttwo}{T$_2$}
\newcommand{\ttwoi}{T$_{2,i}$}
\newcommand{\ttwoo}{T$_{2,o}$}

\newcommand{\tone}{T$_1$}
\newcommand{\ttwoq}{\rm{T}_2}
\newcommand{\toneq}{\rm{T}_1}

\begin{document}

\title{Environmental effects on electron spin relaxation in \natc}

\author{John~J.~L.~Morton}
\email{john.morton@materials.ox.ac.uk} \affiliation{Department of
Materials, Oxford University, Oxford OX1 3PH, United Kingdom}
\affiliation{Clarendon Laboratory,
Department of Physics, Oxford University, Oxford OX1 3PU, United
Kingdom}

\author{Alexei~M.~Tyryshkin}
\affiliation{Department of Electrical Engineering, Princeton
University, Princeton, NJ 08544, USA}

\author{Arzhang~Ardavan}
\affiliation{Department of Materials, Oxford University, Oxford
OX1 3PH, United Kingdom}
\affiliation{Clarendon Laboratory,
Department of Physics, Oxford University, Oxford OX1 3PU, United
Kingdom}

\author{Kyriakos~Porfyrakis}
\affiliation{Department of Materials, Oxford University, Oxford
OX1 3PH, United Kingdom}

\author{S.~A.~Lyon}
\affiliation{Department of Electrical Engineering, Princeton
University, Princeton, NJ 08544, USA}

\author{G.~Andrew~D.~Briggs}
\affiliation{Department of Materials, Oxford University, Oxford
OX1 3PH, United Kingdom}

\date{\today}
% It is always \today, today, but any date may be explicitly specified

\begin{abstract}
We examine environmental effects of surrounding nuclear spins on
the electron spin relaxation of the \natc~molecule (which consists
of a nitrogen atom at the centre of a fullerene cage). Using
dilute solutions of \natc~in regular and deuterated toluene, we
observe and model the effect of translational diffusion of nuclear
spins of the solvent molecules on the \natc~electron spin
relaxation times. We also study spin relaxation in frozen
solutions of \natc~in \cstwo, to which small quantities of a
glassing agent, \stwocltwo~are added. At low temperatures, spin
relaxation is caused by spectral diffusion of surrounding nuclear
$^{35,37}$Cl spins in the \stwocltwo, but nevertheless, at 20~K,
\ttwo~times as long as 0.23~ms are observed.

\end{abstract}


\maketitle

\section{Introduction}

The \natc~molecule is well known for its remarkably
well-shielded electron spin~\cite{knapp97}, prompting several
proposals for fullerene-based quantum information processing
(QIP)~\cite{harneit,frankenstein}. Indeed, extraordinarily long
electron spin relaxation times, satisfying the strict requirements for
QIP~\cite{harneit}, have been reported for \natc~in liquid
solutions~\cite{dietel99,relaxcs2} and in solid
matrices~\cite{harneit}, thus demonstrating the remarkable capacity
of the fullerene cage for protecting the enclosed spin from fluctuating
perturbations in various host environments. This property opens the
possibility of using almost any host material when
``designing" \natc-based QIP processors. For example, in proposed
architectures include \natc~arrays positioned at
interfaces (e.g. arranged on solid templates) where they would be expected to be
exposed to a broad spectrum of environmental
perturbations.

Two important spin relaxation mechanisms have recently been
identified for \natc~in liquid solutions~\cite{knapp97,relaxcs2},
both involving internal motion of the fullerene cage (e.g.
vibrational or rotational motion). These two mechanisms explain a
large body of the experimental data. An Orbach mechanism via a
vibrational mode of \csixty~cage was shown to determine the spin
relaxation of \natc~in a \cstwo~solvent environment over a broad
range of temperatures~\cite{relaxcs2}. Nevertheless, this Orbach
mechanism remains weak resulting in very long relaxation times
(\tone=0.5~ms and \ttwo=0.24~ms at 160~K, just above the melting
point of \cstwo). Even longer relaxation times might be expected
at lower temperatures; however, the \cstwo~solvent is not suitable
for frozen solution studies since it freezes as a polycrystal with consequent grain boundary segregation of the dissolved fullerene molecules.

A second relaxation mechanism was found for asymmetric
\natcseventy~fullerenes, which possess a permanent zero field
splitting (ZFS)~\cite{relaxcs2}. Random rotational reorientation
of this ZFS contributes significantly to \natcseventy~spin
relaxation at temperatures lower than 260~K, a temperature range in which rotational
mobility is insufficient to achieve efficient motional
averaging of the non-zero ZFS.

Relaxation of \natc~(and \natcseventy) in solid matrices has not
been comprehensively studied yet. Very long \tone$\sim 1$~s have
been reported in low-purity \natc/\csixty~powders at 4~K, while
considerably shorter \ttwo$=20 \mu s$ were found~\cite{harneit}.
The mechanism behind such an unexpectedly short \ttwo~remains
unexplained.

% For low
%temperature work, it is therefore necessary to either use a
%solvent which freezes as a glass (such as toluene), or to add a
%glassing agent such as \stwocltwo~(sulphur chloride) to the
%primary \cstwo~solvent. Both options necessarily involve
%introducing a significant nuclear spin concentration to the
%\natc~environment, leading to further potential relaxation
%pathways.

In this paper, we extend the studies of electron spin relaxation
of \natc~and examine the role of nuclear spins in the solvent
environment, both in liquids and in frozen solutions. The
\cstwo~solvent used in our previous studies had no naturally
abundant nuclear spins (i.e.\ only 1.1\% of $^{13}$C with nuclear
spin I=1/2, and 0.76\% of $^{33}$S with I=3/2). In this work we
use a toluene solvent and a \cstwo/\stwocltwo~mixture: both
contain substantial numbers of magnetic nuclei. We show how the
presence of a high concentration of nuclear spins from solvent
molecules can significantly shorten the relaxation time of \natc.
Depending on the temperature regime, i.e.\ liquid or frozen
solutions, translational~\cite{bancibook} or
spectral~\cite{Klauder62,Zhidomirov69,Milov73, nevzorov02}
diffusion of nuclear spins surrounding the \natc~molecule
contributes to electron spin relaxation.

\section{Materials and Methods}

The aggregation (or clustering) of \csixty~in certain solvents and
concentrations has been widely reported~\cite{bokare03, cluster1,
cluster2, cluster5, cluster4, cluster6, cluster7}. Furthermore,
the solubilities of \csixty~in toluene and in \cstwo~show a strong
temperature dependence, peaking at 280~K and falling rapidly upon
further cooling~\cite{cluster9}.  The result is that the
convenient picture of isolated fullerenes in solution is rather
na\"ive; instead the behaviour is a complex, non-monotonic function of
temperature, fullerene concentration, choice of solvent, and even
time from initial dissolution. For example, in \cstwo~at room
temperature the onset of aggregation has been measured to be
at a concentration of around 0.06~mg/ml~\cite{bokare03}. At concentrations above
0.36~mg/ml, the clusters themselves further agglomerate to form
`flowerlike' structures with an open hole in the centre. In
toluene, clusters ranging from 3 to 55 fullerenes have been
observed over a dilute range of concentrations (0.18 to
0.78~mg/ml)~\cite{cluster7}. This clustering can have important
consequences on electron spin relaxation rates resulting in a
distribution of the relaxation times depending on the location of
\natc~within the cluster. An additional complication can arise in
samples of higher \natc/\csixty~purity --- if the large
\csixty~cluster contains two \natc~molecules their
relaxation will be strongly affected by the dipole-dipole
interaction between the two \natc~electron spins. For example, in a
sample of 3$\%$ \natc/\csixty~purity, we have observed a decrease
in \ttwo~with increasing fullerene concentration above about
0.1~mg/ml.

To eliminate uncertainties associated with \csixty~cluster formation,
dilute solutions with concentrations of less than 0.06~mg/ml were
used in this study. High-purity ($\approx80\%$) endohedral
\natc~was used to prepare samples in toluene, enabling the use of
dilute solutions (2~$\mu$g/ml) of well-isolated fullerenes which
nevertheless provide sufficient signal for pulsed EPR experiments.
Solutions were degassed by freeze-pumping in three cycles to
remove paramagnetic O$_2$. We observe that while samples of
\natc~in \cstwo~are stable, the EPR signal from the sample in
degassed toluene decayed when exposed to light. The precise
nature of this decay is unknown and possibly occurs via the
photo-excited triplet state of the \csixty~cage leading to escape
of the nitrogen from the cage. Other experimental parameters,
including a brief description of the \natc~spin system are
provided elsewhere~\cite{relaxcs2}. \ttwo and \tone times were obtained
using Hahn echo and inversion recovery sequences, respectively~\cite{schweiger2001}.

Given the strong reactivity of \stwocltwo, there was some concern
that it might attack the fullerenes in the solution --- this
dictated the sample preparation procedure adopted. 50~$\mu$l
samples of \natc~(4$\%$ purity)~in \cstwo, and pure
\stwocltwo~were degassed in two separate arms of a
$\lambda$-shaped quartz vessel. The solvents were then mixed and
quickly frozen; the resulting mixture contained approximately
25$\%$ \stwocltwo~by volume (corresponding to 20~mol$\%$).

%\section{Fullerene aggregation effects} \label{sec:aggre}
%
%\begin{figure}[t] \centerline
%{\includegraphics[width=2.3in]{Figs/cluster.eps}} \caption{Decoherence
%times of \natc~in \cstwo~solutions as a function of concentration, measured at 300K.} \label{cluster}
%\end{figure}
%
%\Fig{cluster}
%
%Samples of a lower \natc/\csixty~purity (0.1$\%$) showed no
%pronounced dependence on concentration, as dipolar interactions
%within such (\csixty)$_n$ clusters is much weaker.

%There is a large parameter space to be investigated in order to further map out
%the effects of fullerene aggregation on the relaxation properties of \natc.
%Such a study is beyond the intentions of this report, but these results confirm
%that it is indeed an effect which must be considered and which merits further attention.

\section{Spin relaxation of \natc~in toluene solution} \label{sec:tolrelax}
\begin{figure}[t] \centerline
{\includegraphics[width=3.3in]{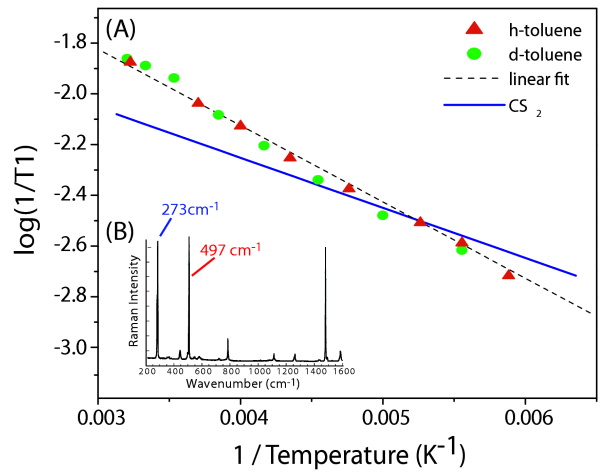}} \caption{(A) Spin
relaxation time \tone~for \natc~in toluene as a function of
temperature, plotted as $\log{(1/\toneq)}$ against $(1/T)$, where
\tone~is given in microseconds. \tone~times are indistinguishable
for regular toluene (red triangles) and deuterated toluene
(green circles). The linear fit (dashed line) is consistent with
an Orbach relaxation mechanism, and the slope to the fit gives an
energy splitting $\Delta = 60(2)$~meV of the excited state
involved in the relaxation process. Similar linear dependence has
been reported for \natc~in \cstwo~\cite{relaxcs2}, but the linear
fit (shown in blue, for comparison) gave a different $\Delta =
33$~meV. (B) Two major absorption peaks in the Raman spectrum of
\csixty~lie at 273 and 497~cm$^{-1}$ (33 and 62~meV,
respectively).} \label{tolt1}
\end{figure}

\begin{figure}[t] \centerline
{\includegraphics[width=3.5in]{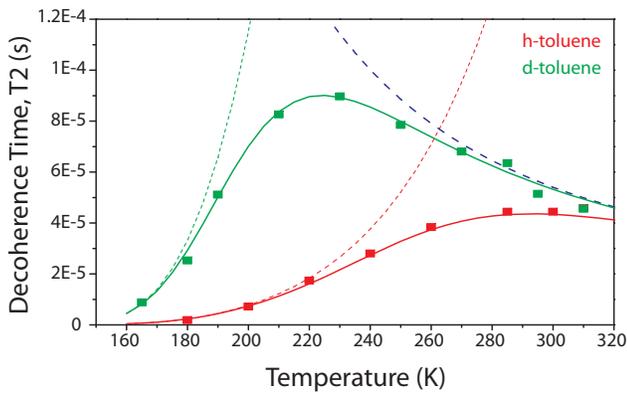}} \caption{Spin
decoherence time \ttwo~of \natc~in toluene, measured using the
central $M_I=0$ line, with regular toluene (red squares) and
deuterated toluene (green squares). The solid curves are generated
using the model that involves two relaxation mechanisms, and the
dashed curves show individual contributions of the two mechanisms.
The blue dashed curve is the `intrinsic' decoherence time due to the
Orbach relaxation process; \ttwo = (2/3)\tone~is assumed based on
the study of \natc~in \cstwo~\cite{relaxcs2}. The red and green
dashed curves show the relaxation effect due to translational
diffusion of proton and deuterium nuclei of toluene molecules.}
\label{tolrelax}
\end{figure}

When \natc~is dissolved in a solution containing nuclear spins
(such as the hydrogen atoms of toluene), additional relaxation
pathways may be introduced.  These arise from fluctuating fields
caused by the motion of solvent molecules around the fullerene
cage. The use of both regular (hydrogenated) and deuterated
toluene as solvents can provide further insights into the effect
of local nuclear spins. Figures~\ref{tolt1} and
\ref{tolrelax}~show the \tone~and \ttwo~times measured for
high-purity \natc~in toluene solution, as a function of
temperature.

The temperature dependence of \tone~in toluene is suggestive of an
Orbach relaxation mechanism, similar to that reported for \natc~in
\cstwo~\cite{relaxcs2}. However, the slopes of the temperature
dependence are markedly different in the two solvents. In \cstwo~the
energy splitting derived from the slope matched well the first
excited H$_g$(1) vibrational mode of \csixty~at 33~meV, and in
toluene the slope corresponds to an energy splitting of 60(2)~meV
which coincides with the second major line seen in the Raman
spectra (62~meV), corresponding to the A$_g$(1) mode
(\Fig{tolt1}B). Solvent effects have been reported extensively in
the Raman spectroscopy of \csixty~\cite{gallagher96}, and it is
concluded that the nature of the solvent-fullerene interaction can
distort the icosahedral symmetry leading to splittings of the
H$_g$ Raman transitions~\cite{gallagher97}. Consistently, the
results here could also be attributed to interactions between the
cage and the solvent (e.g. a $\pi$-stacking arrangement in the
case of the aromatic toluene molecule); the transitions involving
the H$_g$(1) mode may be suppressed, and electron spin relaxation
of the endohedral nitrogen takes places more effectively via the
higher-energy A$_g$(1) squeezing mode.

The \ttwo~relaxation data in \Fig{tolrelax} reveal a non-monotonic
temperature dependence in contrast to that observed for \natc~in
\cstwo~\cite{relaxcs2}. In \cstwo, a simple ratio of
$\ttwoq=2/3~\toneq$ was found over this broad temperature range
indicating that both \tone~and \ttwo~times are determined by the
same Orbach relaxation mechanism. In toluene, \ttwo~diverges
noticeably from the \tone~dependence indicating that an additional
relaxation mechanism must be involved which suppresses \ttwo~at
low temperatures. In the following discussion we argue that this
additional relaxation mechanism is due to nuclear spins (protons)
of the toluene solvent.

In liquid solutions, solvent molecules can diffuse around
\natc~and therefore the distance between the electron spin of
\natc~and the nuclear spins of toluene molecules changes in time.
This results in fluctuating hyperfine (contact and dipolar) fields
seen by the electron spin which can drive its relaxation. In the
hard-sphere approximation, the spin-spin separation varies between
a value called the \emph{distance of closest approach} ($d$), and
infinity. The translation diffusion time, $\tau_D$, becomes the
important correlation time~\cite{bancibook},
\beq \tau_D=\frac{2d^2}{D(T)}, \eeq %
where $D(T)=D_{\mathrm{C_{60}}}(T)+D_{\mathrm{tol}}(T)$ is the sum of the
temperature-dependent diffusion coefficients of the fullerene and
toluene molecules. According to common models for
diffusion-induced spin
relaxation~\cite{hwang75,freed78,polnaszek84}, the resulting
\tone~and \ttwo~times are~\cite{bancibook}

\beq\label{difft1} ({\rm T}_1)^{-1}=2\kappa~\frac{c(T)}{d \cdot
D(T)}~10 J\left(\omega_e\right),\eeq

\beq \label{difft2}
({\rm T}_2)^{-1}=\kappa~\frac{c(T)}{d \cdot D(T)}~\left[4J\left(0\right)+10J\left(\omega_e\right)+6J\left(\omega_n\right)\right].\eeq %
$\omega_e$ and $\omega_n$ are the electron and nuclear Zeeman
frequencies, respectively. The constant prefactor, $\kappa$, is
given by

\beq \kappa=\frac{16\pi}{405}\gamma_e^2 \gamma_n^2 \hbar^2~I(I+1).
\eeq %
$\gamma_e$ and $\gamma_n$ are the electron and nuclear
gyromagnetic ratios, $c(T)$ is the temperature-dependent
concentration of hydrogen (or deuterium) spins, and the spectral
density function $J(\omega)$ is given by

\beq
J(\omega)=\frac{1+5z/8+z^2/8}{1+z+z^2+z^3/6+4z^4/81+z^5/81+z^6/648},
\eeq with $z = \sqrt{2 \omega \tau_D}$.

The only unknown quantities in the above expressions are
nuclear spin concentration $c(T)$, distance of closest approach
$d$, and the diffusion coefficient $D(T)$. The temperature
dependence of $^1$H concentration in toluene is given in Perry's
Chemical Engineer's Handbook (7th Ed) as

\beq
c(T)=\left(4.089\cdot10^{21}\right)\left(0.26655^{-\left(1+\left(1-\frac{T}{591.8}\right)^{0.2878}
\right)}\right), \eeq %
and varies between about $4.5 - 5\cdot10^{22}$~cm$^{-3}$ over the
temperature range $150 - 300$K. Evidently, the variation in this
parameter is not great and therefore could not explain the
temperature dependence of \ttwo. It must be $D(T)$ and its strong
temperature dependence that dominates the effect on \ttwo.

The self-diffusion coefficient $D(T)$ of toluene has been studied
as a function of temperature~\cite{selfdiff1}. In the temperature
range 135 to 330K, the data fit well to

\beq \label{Dtol} D_{\mathrm{tol}}(T)=
6.1\cdot10^{-4}~\exp{\left(-\frac{1000}{T}\right)}~\exp{\left(-\left(\frac{190}{T}\right)^6\right)}
\eeq %

The diffusion coefficient for \csixty~can be roughly estimated by
the Stokes-Einstein equation:

\beq \label{viscdiff} D_{\mathrm{C_{60}}}(T)=\frac{k_BT}{6\pi
a\eta(T)},\eeq %
where $a = 0.35$~nm is the radius of the \natc~molecule, and
$\eta(T)$ is the solvent viscosity which can also be temperature
dependent. However, reports on toluene viscosity only go down to
225K~\cite{tolvisc}, below which one would expect substantial
changes in behaviour. As a result, both $D_{\mathrm{C_{60}}}(T)$ and $d$
were left as fitting parameters.

\begin{figure}[t] \centerline
{\includegraphics[width=3.3in]{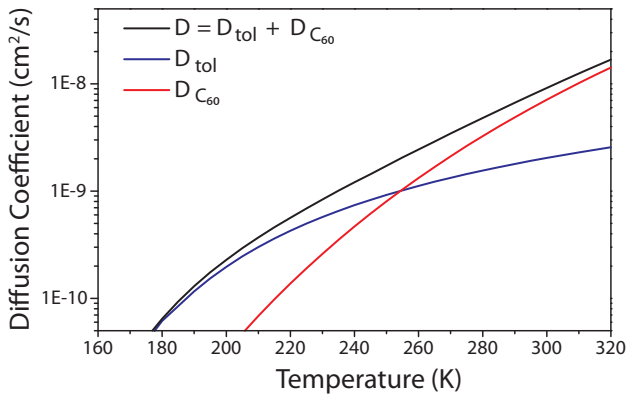}}
\caption{Temperature dependence for (blue curve) self-diffusion
coefficient of toluene, $D_{\mathrm{tol}}$, based on experimental
results~\cite{selfdiff1}; (red curve) the predicted diffusion
coefficient of \csixty~in toluene, $D_{\mathrm{C_{60}}}$, which produces
the best fits to the data in \Fig{tolrelax}; (black curve) the
overall diffusion coefficient, $D = D_{\mathrm{tol}} + D_{\mathrm{C_{60}}}$.}
\label{diffcoefs}
\end{figure}

To fit the experimental data in \Fig{tolrelax} we assume two
relaxation processes: the translational diffusion mechanism
described above and the Orbach relaxation mechanism which if alone
would produce \ttwo~=~(2/3)~\tone, as was found for \natc~in
\cstwo~solution~\cite{relaxcs2}. The individual contribution of
each of the two relaxation mechanisms and their overall effect are
shown in \Fig{tolrelax}. The best fit was achieved using a
diffusion coefficient whose temperature dependence is shown in
\Fig{diffcoefs}, and $d=0.35$~nm (though it was possible to obtain
reasonable fits for $d$ up to about 0.45~nm). The radius of the
\csixty~molecule is 0.35~nm, so these values for distance of
closest approach are reasonable. The best-fit diffusion
coefficient of \csixty~converges with that predicted by the
Stokes-Einstein equation (Eq.~\ref{viscdiff}) for temperatures
below 250K, however it deviates by as much as a factor of 10 at
higher temperatures (310K).

\begin{figure}[t] \centerline
{\includegraphics[width=3.3in]{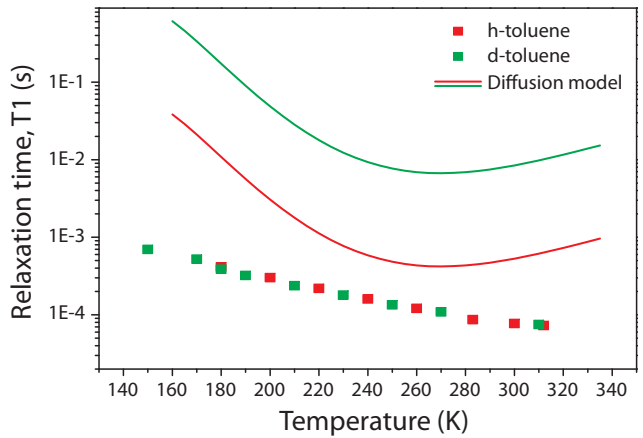}}
\caption{(Squares) Experimental \tone~values for \natc~in regular
(red) and deuterated (green) toluene. (Curves) Theoretical
\tone~predicted from the model of relaxation by translational
diffusion of nuclear spins in the solvent molecules
(Eq.~\ref{difft1}).} \label{tolt1model}
\end{figure}

Finally, evaluating Eq.~\ref{difft1} with the parameters extracted
from the study of \ttwo, we confirm that the \tone~times for
h-toluene and d-toluene are expected to be equal, as the
translational diffusion \tone~rates are much slower than the
intrinsic (Orbach) decay mechanism (see \Fig{tolt1model}).

\section{Spin relaxation of \natc~in a frozen solution}

A glass-forming solvent is essential for frozen solution studies
of \natc, in order to ensure homogeneity of the frozen solution
and to avoid clustering of \natc. The ideal solvent would also
contain a minimal concentration of nuclear spins since it is known
that nuclear spins of the solvent molecules can provide a
significant mechanism for electron spin decoherence, e.g.\ via the
process known as \emph{spectral diffusion} caused by flip-flops of
the local nuclear spins~\cite{Klauder62,Zhidomirov69,Milov73}.
While such an ideal solvent has not come to our attention, it is
possible to add relatively small quantities of sulphur chloride
(\stwocltwo) to \cstwo~to act as a glassing agent~\cite{cs2glass}.
The addition of 15~mol$\%$ \stwocltwo~in \cstwo~is sufficient to
permit vitrification of small samples. \cstwo~has no major
isotopes with non-zero nuclear spins, however, \stwocltwo~has
chlorine whose major isotopes both have nuclear spin $I=3/2$ and
gyromagnetic ratios of about 4~MHz/T. Therefore, while this
mixture is not an optimal solution, it was hoped that the reduced
nuclear spin concentration in the mixture solution would permit
relatively long decoherence times, not limited by the nuclear
spectral diffusion.

\begin{figure}[t] \centerline
{\includegraphics[width=3.3in]{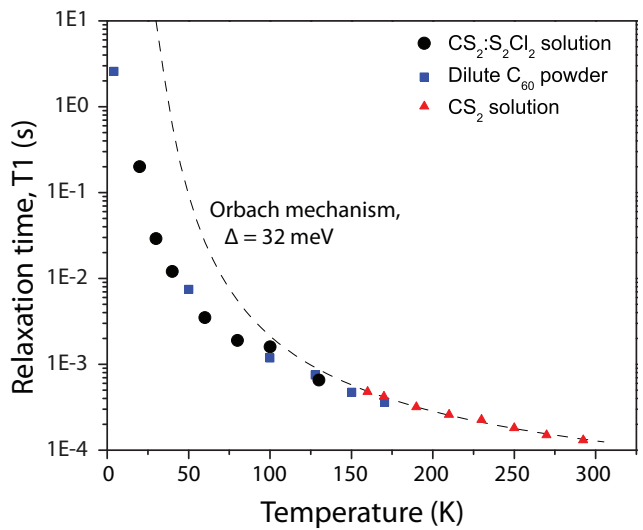}}
\caption{Temperature dependence of \tone~for \natc~in a frozen
solution of \cstwo~:~\stwocltwo~(volume 3:1), measured on the
central ($M_I=0$) hyperfine line. Data for \natc~diluted in
\csixty~powder, and in \cstwo~solution are shown for comparison.}
\label{s2cl2t1}
\end{figure}

\begin{figure}[t] \centerline
{\includegraphics[width=3.3in]{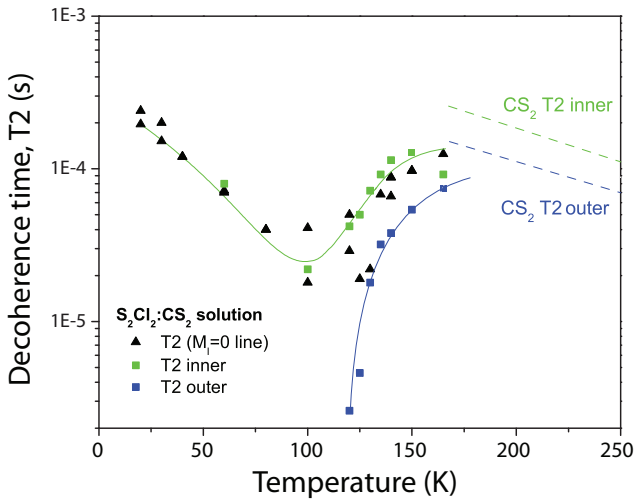}}
\caption{Temperature dependence of \ttwo~for
\natc~in a frozen solution of \cstwo~:~\stwocltwo~(volume 3:1),
measured on the central ($M_I=0$) and high-field ($M_I=-1$)
hyperfine lines of the \natc~EPR spectrum. For $M_I=-1$, two
\ttwo~times, \ttwo$_o$ and \ttwo$_i$, are extracted for outer
($M_S=\pm3/2:\pm1/2$) and inner ($M_S=+1/2:-1/2$) transitions
within the S=3/2 multiplet, using their different ESEEM
frequencies~\cite{relaxcs2}. The curves are visual guides. }
\label{s2cl2t2}
\end{figure}

The measured \tone~and \ttwo~times are shown in Figures \ref{s2cl2t1}
and \ref{s2cl2t2} in the temperature range 20 to 165K (below the
melting point of the mixture). \tone, which was measured on the
central $M_I=0$ hyperfine line, has a monotonic temperature
dependence and follows closely that seen in a dilute powder of
\natc/\csixty. It seems that the residual concentration of nuclear
spins in \stwocltwo/\cstwo~mixture has no effect on \tone~in the
temperature range studied. However, both solid samples show
\tone~values which are less than that expected by extrapolating
the Orbach mechanism suggested for \natc~in \cstwo~solution.
Apparently, other (yet unidentified) relaxation mechanisms
contribute significantly in both solid matrices at low
temperatures, resulting in \tone~shorter than would be expected
from the Orbach mechanism alone. Relaxation experiments at
different microwave frequencies will be required to shed light on
these unidentified mechanisms.

On the other hand, \ttwo, when measured on the $M_I=0$ line, shows
a minimum at around 100~K, coinciding with the approximate glass
transition temperature, $T_g$, of the solvent
mixture~\cite{cs2glass}. For measurements on the $M_I=-1$
high-field hyperfine line, two different \ttwo~times corresponding
to the outer ($M_S=\pm3/2:\pm1/2$) and inner
($M_S=+1/2:-1/2$) transitions can be separated using the ESEEM
method described elsewhere~\cite{relaxcs2}. \ttwoo~(outer) falls
dramatically upon cooling towards $T_g$. \ttwoi~(inner) reaches a
minimum at $T_g$, but then rises as the temperature is lowered
further. We can now see that the \ttwo~measured on the central
line is indeed a weighted average of the two separate \ttwo~times
(inner and outer), for temperatures above $T_g$. Below $T_g$, the
outer coherences decay sufficiently quickly (\ttwoo$<1\mu$s)
and thus they become unobservable in spin echo experiment. Indeed,
only \ttwoi~can be measured below $T_g$. Recognizing the fact
that \ttwoi~measured for $M_I=0$ and $-1$ are almost identical
(within the error of the experiment) in a broad temperature range
below 120K, we can conclude that hyperfine interaction with
$^{14}$N nucleus has no visible effect on \ttwo~in \natc~at low
temperatures.

The fact that \ttwoo~is much shorter than \ttwoi~at low temperatures can
be explained by the effect of a non-zero ZFS. The glassy solvent matrix
around the fullerene imposes a distortion of the \natc~cage,
inducing a ZFS strain to the nitrogen atom. At temperatures
approaching $T_g$ and below, the rotational mobility of fullerenes
slows down and therefore complete motional averaging of the
non-zero ZFS is not achieved. In this case, the ZFS strain creates
an inhomogeneous broadening of the EPR transitions, and the outer
transitions within the S=3/2 multiplet are broadened more
significantly, i.e. via the first-order ZFS, in contrast to the inner
transition which are broadened only via the second-order ZFS. Provided the
ZFS broadening is small compared to the excitation bandwidth of
the microwave pulses, and thus the second pulse in spin echo
experiment refocuses all transitions within the S=3/2 multiplet,
such a ZFS broadening would not be refocused for the outer
coherences and appears instead as an additional dephasing
mechanism, leading to fast decay of the outer transitions.
On the other hand, the inner transition is fully refocused, and
its decoherence time, \ttwoi, appears unaffected by the ZFS.

As follows from the above discussion, the two strongest
interactions in \natc, the ZFS strain and the hyperfine coupling
to the central $^{14}$N nucleus, have little effect on \ttwoi.
Therefore, some other interactions need to be considered to
explain \ttwoi~and its temperature dependence. Here we propose
that these mechanisms involve nuclear $^{35}$Cl and $^{37}$Cl
spins of solvent molecules in frozen \stwocltwo/\cstwo~mixture.
The theory of decoherence for electron spins interacting with a
bath of nuclear spins (so-called theory of \emph{spin diffusion})
has been developed decades
ago~\cite{Klauder62,Zhidomirov69,Milov73}. More recently the
theory has been extended beyond stochastic treatment of random
nuclear flip-flops to a coherent treatment of all spins together
as a single many-body system~\cite{nevzorov01,nevzorov02}. The new
theory is also robust to translational diffusion of solvent
molecules, covering the fast and slow diffusion regimes, from
motional narrowing at high temperatures to the rigid limit at low
temperatures. Thus this theory is ideal for the temperature range
studied here. We notice however that this theory has been
developed for a bath of nuclear spins I=1/2 and as such it may only be
used as an approximation in analyzing our data for I=3/2
of $^{35}$Cl and $^{37}$Cl.

In the rigid limit, i.e.\ no translational diffusion, spin
diffusion theory predicts a stretched exponential dependence of
the echo decays, e.g. $V(\tau) = A \cdot \exp(-(2\tau/T_{2})^n)$,
with $2\le n \le 3$, and $\tau$ being the delay between pulses in a
two-pulse echo
experiment~\cite{Klauder62,Zhidomirov69,Milov73,nevzorov01}. This
dependence seems to be inconsistent with the simple exponential
decay, $V(\tau) = A \cdot \exp(-2\tau/T_{2,i})$, observed in our
experiments for \natc~in \stwocltwo/\cstwo~at all temperatures
studied from 20 K to 170 K.

An echo decay close to a simple exponential is predicted in the
slow diffusion regime and assuming low nuclear spin
concentrations~\cite{nevzorov02}. In this case the echo decay is
dominated by a term $\tau^{9/8}$ in the exponential, which is
close enough to be indistinguishable from a simple exponential
decay in our experiments. For the \cstwo/\stwocltwo~(3/1) mixture used
here, we can estimate $c\cdot d^3 < 0.1$ (where $c$ is
concentration of $^{35}$Cl or $^{37}$Cl nuclei, and $d$ is
distance of closest approach between electron and nuclear spins)
which satisfies the derived criteria for low nuclear spin
concentration~\cite{nevzorov02}. Using Eqn.(3.7) from
Ref.~\cite{nevzorov02}, we can reproduce the mono-exponential
decay with \ttwoi$=230~\mu$s for \natc~at 20~K assuming a
diffusion coefficient $D=5\cdot 10^{-16}$~cm$^2$/s for solvent
molecules in \cstwo/\stwocltwo. This $D$ is small but nevertheless large
enough for the exponential term $\tau^{9/8}$ to dominate over spin
diffusion term $\tau^{2}$ expected in the rigid
limit~\cite{nevzorov01}.

As temperature increases from 20~K (and consequently $D$ also
increases), \ttwo~is predicted to initially decrease and then
increase after reaching a minimum at temperature where $D \approx
0.1 \gamma_e \gamma_n \hbar \cdot d$~\cite{nevzorov01}. This
minimum corresponds to a transition from the intermediate to fast
diffusion regimes; a mono-exponential term continues to dominate
spin echo decay in both these regimes. For \cstwo/\stwocltwo~this
$T_{2}$ minimum is expected to occur at temperature where $D =
10^{-10}~$cm$^2$/s. Comparing with our \ttwoi~data in
Figure~\ref{s2cl2t2}, we see that \ttwoi~indeed develops a minimum
at around 100~K. Thus, as temperature increases from 20 to 100~K,
$T_{2,i}$ decreases by one order of magnitude (from 230~$\mu$s to
20~$\mu$s) and $D$ increases by five orders of magnitude (from
$10^{-15}~$cm$^2$/s to $10^{-10}~$cm$^2$/s). This corresponds to
an approximate dependence \ttwo~$\sim D^{-0.2}$ and thus differs
slightly from $D^{-0.36}$ predicted by the theory for
I=1/2~\cite{nevzorov01}. As temperature increases beyond
100~K, \ttwoi~increases to 100~$\mu$s at 165~K, and we can infer that
$D$ increases only by about an order of magnitude.

To conclude, by comparing our data
with the relationship between decoherence rate and diffusion
coefficient described in Ref.~\cite{nevzorov01}, we infer that $D$
changes relatively slowly above about 100~K, in contrast to the
sharp drop in $D$ observed below 100~K. This is consistent with
the expected behaviour around the glass transition temperature,
T$_g$.

\section{Conclusions}
In summary, the effect of nuclear spins in toluene surrounding
\natc~on the decoherence time has been demonstrated using two
different isotopes of hydrogen. The data fit well to a model
for relaxation by translational diffusion, providing estimates of
the diffusion coefficient for \csixty~in toluene.

The nuclear spin concentration can be reduced by using a solvent
mixture of \cstwo~and \stwocltwo. Using such mixture, decoherence
times approaching 0.23~ms were observed at temperatures below 20K,
demonstrating this to be a good choice of solvent for
low-temperature studies on \natc. Below about 100~K, the outer
coherences are not refocussed, turning \natc~into a `quasi' S=1/2
spin system. Translational diffusion of local nuclear spins
continues to play the dominant role in decoherence, even for an
estimated $D$ as low as $10^{-15}~$cm$^2$/s.

\section{Acknowledgements}
We acknowledge helpful discussions with Mark Sambrook, and thank
the Oxford-Princeton Link fund for support. This research is part
of the QIP IRC www.qipirc.org (GR/S82176/01) and was supported by EPSRC grant number EP/D048559/1. JJLM is supported by
St. John's College, Oxford. AA is supported by the Royal Society.
GADB is supported by the EPSRC (GR/S15808/01). Work at Princeton
was supported by the NSF International Office through the
Princeton MRSEC Grant No. DMR-0213706 and by the ARO and ARDA
under Contract No. DAAD19-02-1-0040.

%\bibliography{bib}

\begin{thebibliography}{28}
\expandafter\ifx\csname natexlab\endcsname\relax\def\natexlab#1{#1}\fi
\expandafter\ifx\csname bibnamefont\endcsname\relax
  \def\bibnamefont#1{#1}\fi
\expandafter\ifx\csname bibfnamefont\endcsname\relax
  \def\bibfnamefont#1{#1}\fi
\expandafter\ifx\csname citenamefont\endcsname\relax
  \def\citenamefont#1{#1}\fi
\expandafter\ifx\csname url\endcsname\relax
  \def\url#1{\texttt{#1}}\fi
\expandafter\ifx\csname urlprefix\endcsname\relax\def\urlprefix{URL }\fi
\providecommand{\bibinfo}[2]{#2}
\providecommand{\eprint}[2][]{\url{#2}}

\bibitem[{\citenamefont{Knapp et~al.}(1997)\citenamefont{Knapp, Dinse, Pietzak,
  Waiblinger, and Weidinger}}]{knapp97}
\bibinfo{author}{\bibfnamefont{C.}~\bibnamefont{Knapp}},
  \bibinfo{author}{\bibfnamefont{K.-P.} \bibnamefont{Dinse}},
  \bibinfo{author}{\bibfnamefont{B.}~\bibnamefont{Pietzak}},
  \bibinfo{author}{\bibfnamefont{M.}~\bibnamefont{Waiblinger}},
  \bibnamefont{and}
  \bibinfo{author}{\bibfnamefont{A.}~\bibnamefont{Weidinger}},
  \bibinfo{journal}{\CPL} \textbf{\bibinfo{volume}{272}}, \bibinfo{pages}{433}
  (\bibinfo{year}{1997}).

\bibitem[{\citenamefont{Harneit}(2002)}]{harneit}
\bibinfo{author}{\bibfnamefont{W.}~\bibnamefont{Harneit}},
  \bibinfo{journal}{\PRA} \textbf{\bibinfo{volume}{65}}, \bibinfo{pages}{32322}
  (\bibinfo{year}{2002}).

\bibitem[{\citenamefont{Benjamin et~al.}((2004))\citenamefont{Benjamin,
  Ardavan, Briggs, Britz, Gunlycke, Jefferson, Jones, Leigh, Lovett, Khlobystov
  et~al.}}]{frankenstein}
\bibinfo{author}{\bibfnamefont{S.~C.} \bibnamefont{Benjamin}},
  \bibinfo{author}{\bibfnamefont{A.}~\bibnamefont{Ardavan}},
  \bibinfo{author}{\bibfnamefont{G.~A.~D.} \bibnamefont{Briggs}},
  \bibinfo{author}{\bibfnamefont{D.~A.} \bibnamefont{Britz}},
  \bibinfo{author}{\bibfnamefont{D.}~\bibnamefont{Gunlycke}},
  \bibinfo{author}{\bibfnamefont{J.}~\bibnamefont{Jefferson}},
  \bibinfo{author}{\bibfnamefont{M.~A.~G.} \bibnamefont{Jones}},
  \bibinfo{author}{\bibfnamefont{D.~F.} \bibnamefont{Leigh}},
  \bibinfo{author}{\bibfnamefont{B.~W.} \bibnamefont{Lovett}},
  \bibinfo{author}{\bibfnamefont{A.~N.} \bibnamefont{Khlobystov}},
  \bibnamefont{et~al.}, \emph{\bibinfo{title}{quant-ph/0511198}}
  (\bibinfo{year}{(2004)}).

\bibitem[{\citenamefont{Dietel et~al.}(1999)\citenamefont{Dietel, Hirsch,
  Pietzak, Waiblinger, Lips, Weidinger, Gruss, and Dinse}}]{dietel99}
\bibinfo{author}{\bibfnamefont{E.}~\bibnamefont{Dietel}},
  \bibinfo{author}{\bibfnamefont{A.}~\bibnamefont{Hirsch}},
  \bibinfo{author}{\bibfnamefont{B.}~\bibnamefont{Pietzak}},
  \bibinfo{author}{\bibfnamefont{M.}~\bibnamefont{Waiblinger}},
  \bibinfo{author}{\bibfnamefont{K.}~\bibnamefont{Lips}},
  \bibinfo{author}{\bibfnamefont{A.}~\bibnamefont{Weidinger}},
  \bibinfo{author}{\bibfnamefont{A.}~\bibnamefont{Gruss}}, \bibnamefont{and}
  \bibinfo{author}{\bibfnamefont{K.-P.} \bibnamefont{Dinse}},
  \bibinfo{journal}{\JACS} \textbf{\bibinfo{volume}{121}},
  \bibinfo{pages}{2432} (\bibinfo{year}{1999}).

\bibitem[{\citenamefont{Morton et~al.}(2006)\citenamefont{Morton, Tyryshkin,
  Ardavan, Porfyrakis, Lyon, and Briggs}}]{relaxcs2}
\bibinfo{author}{\bibfnamefont{J.~J.~L.} \bibnamefont{Morton}},
  \bibinfo{author}{\bibfnamefont{A.~M.} \bibnamefont{Tyryshkin}},
  \bibinfo{author}{\bibfnamefont{A.}~\bibnamefont{Ardavan}},
  \bibinfo{author}{\bibfnamefont{K.}~\bibnamefont{Porfyrakis}},
  \bibinfo{author}{\bibfnamefont{S.~A.} \bibnamefont{Lyon}}, \bibnamefont{and}
  \bibinfo{author}{\bibfnamefont{G.~A.~D.} \bibnamefont{Briggs}},
  \bibinfo{journal}{\JCP} \textbf{\bibinfo{volume}{124}},
  \bibinfo{pages}{014508} (\bibinfo{year}{2006}).

\bibitem[{\citenamefont{Banci et~al.}(1991)\citenamefont{Banci, Bertini, and
  Luchinat}}]{bancibook}
\bibinfo{author}{\bibfnamefont{L.}~\bibnamefont{Banci}},
  \bibinfo{author}{\bibfnamefont{I.}~\bibnamefont{Bertini}}, \bibnamefont{and}
  \bibinfo{author}{\bibfnamefont{C.}~\bibnamefont{Luchinat}},
  \emph{\bibinfo{title}{Nuclear and Electron Relaxation}}
  (\bibinfo{publisher}{VCH, Weinheim}, \bibinfo{year}{1991}).

\bibitem[{\citenamefont{Klauder and Anderson}(1962)}]{Klauder62}
\bibinfo{author}{\bibfnamefont{J.~R.} \bibnamefont{Klauder}} \bibnamefont{and}
  \bibinfo{author}{\bibfnamefont{P.~W.} \bibnamefont{Anderson}},
  \bibinfo{journal}{\PR} \textbf{\bibinfo{volume}{125}}, \bibinfo{pages}{912}
  (\bibinfo{year}{1962}).

\bibitem[{\citenamefont{Zhidomirov and Salikhov}(1969)}]{Zhidomirov69}
\bibinfo{author}{\bibfnamefont{G.~M.} \bibnamefont{Zhidomirov}}
  \bibnamefont{and} \bibinfo{author}{\bibfnamefont{K.~M.}
  \bibnamefont{Salikhov}}, \bibinfo{journal}{Soviet Physics JETP - USSR}
  \textbf{\bibinfo{volume}{29}}, \bibinfo{pages}{1037} (\bibinfo{year}{1969}).

\bibitem[{\citenamefont{Milov et~al.}(1973)\citenamefont{Milov, Salikhov, and
  Tsvetkov}}]{Milov73}
\bibinfo{author}{\bibfnamefont{A.~D.} \bibnamefont{Milov}},
  \bibinfo{author}{\bibfnamefont{K.~M.} \bibnamefont{Salikhov}},
  \bibnamefont{and} \bibinfo{author}{\bibfnamefont{Y.~D.}
  \bibnamefont{Tsvetkov}}, \bibinfo{journal}{Fiz. Tverd. Tela}
  \textbf{\bibinfo{volume}{15}}, \bibinfo{pages}{1187} (\bibinfo{year}{1973}).

\bibitem[{\citenamefont{Nevzorov and Freed}(2002)}]{nevzorov02}
\bibinfo{author}{\bibfnamefont{A.~A.} \bibnamefont{Nevzorov}} \bibnamefont{and}
  \bibinfo{author}{\bibfnamefont{J.~H.} \bibnamefont{Freed}},
  \bibinfo{journal}{\JCP} \textbf{\bibinfo{volume}{117}}, \bibinfo{pages}{282}
  (\bibinfo{year}{2002}).

\bibitem[{\citenamefont{Bokare and Patnaik}(2003)}]{bokare03}
\bibinfo{author}{\bibfnamefont{A.~D.} \bibnamefont{Bokare}} \bibnamefont{and}
  \bibinfo{author}{\bibfnamefont{A.}~\bibnamefont{Patnaik}},
  \bibinfo{journal}{\JCP} \textbf{\bibinfo{volume}{119}}, \bibinfo{pages}{4529}
  (\bibinfo{year}{2003}).

\bibitem[{\citenamefont{Ying et~al.}(1994)\citenamefont{Ying, Marecek, and
  Chu}}]{cluster1}
\bibinfo{author}{\bibfnamefont{Q.}~\bibnamefont{Ying}},
  \bibinfo{author}{\bibfnamefont{J.}~\bibnamefont{Marecek}}, \bibnamefont{and}
  \bibinfo{author}{\bibfnamefont{B.}~\bibnamefont{Chu}},
  \bibinfo{journal}{\JCP} \textbf{\bibinfo{volume}{101}}, \bibinfo{pages}{2665}
  (\bibinfo{year}{1994}).

\bibitem[{\citenamefont{Bezmel'nitsyn et~al.}(1994)\citenamefont{Bezmel'nitsyn,
  Eletskii, and Sepanov}}]{cluster2}
\bibinfo{author}{\bibfnamefont{V.~N.} \bibnamefont{Bezmel'nitsyn}},
  \bibinfo{author}{\bibfnamefont{A.~V.} \bibnamefont{Eletskii}},
  \bibnamefont{and} \bibinfo{author}{\bibfnamefont{E.~V.~J.}
  \bibnamefont{Sepanov}}, \bibinfo{journal}{\JPC}
  \textbf{\bibinfo{volume}{98}}, \bibinfo{pages}{6665} (\bibinfo{year}{1994}).

\bibitem[{\citenamefont{Nath et~al.}(1998)\citenamefont{Nath, Pal, Palit,
  Sapre, and Mittal}}]{cluster5}
\bibinfo{author}{\bibfnamefont{S.}~\bibnamefont{Nath}},
  \bibinfo{author}{\bibfnamefont{H.}~\bibnamefont{Pal}},
  \bibinfo{author}{\bibfnamefont{D.~K.} \bibnamefont{Palit}},
  \bibinfo{author}{\bibfnamefont{A.~V.} \bibnamefont{Sapre}}, \bibnamefont{and}
  \bibinfo{author}{\bibfnamefont{J.~P.} \bibnamefont{Mittal}},
  \bibinfo{journal}{\JPC~B} \textbf{\bibinfo{volume}{102}},
  \bibinfo{pages}{10158} (\bibinfo{year}{1998}).

\bibitem[{\citenamefont{Alargova et~al.}(2001)\citenamefont{Alargova, Deluchi,
  and Tsujii}}]{cluster4}
\bibinfo{author}{\bibfnamefont{R.~G.} \bibnamefont{Alargova}},
  \bibinfo{author}{\bibfnamefont{S.}~\bibnamefont{Deluchi}}, \bibnamefont{and}
  \bibinfo{author}{\bibfnamefont{K.}~\bibnamefont{Tsujii}},
  \bibinfo{journal}{\JACS} \textbf{\bibinfo{volume}{123}}, \bibinfo{pages}{10}
  (\bibinfo{year}{2001}).

\bibitem[{\citenamefont{Bulavin et~al.}(2001)\citenamefont{Bulavin, Adamenko,
  Yashchuk, Ogul'chansky, Prylutskyy, Durov, and Scarff}}]{cluster6}
\bibinfo{author}{\bibfnamefont{L.~A.} \bibnamefont{Bulavin}},
  \bibinfo{author}{\bibfnamefont{I.~I.} \bibnamefont{Adamenko}},
  \bibinfo{author}{\bibfnamefont{V.~M.} \bibnamefont{Yashchuk}},
  \bibinfo{author}{\bibfnamefont{T.}~\bibnamefont{Ogul'chansky}},
  \bibinfo{author}{\bibfnamefont{Y.~I.} \bibnamefont{Prylutskyy}},
  \bibinfo{author}{\bibfnamefont{S.~S.} \bibnamefont{Durov}}, \bibnamefont{and}
  \bibinfo{author}{\bibfnamefont{P.}~\bibnamefont{Scarff}},
  \bibinfo{journal}{\JMOLLIQ} \textbf{\bibinfo{volume}{93}},
  \bibinfo{pages}{187} (\bibinfo{year}{2001}).

\bibitem[{\citenamefont{Nath et~al.}(2000)\citenamefont{Nath, Pal, and
  Sapre}}]{cluster7}
\bibinfo{author}{\bibfnamefont{S.}~\bibnamefont{Nath}},
  \bibinfo{author}{\bibfnamefont{H.}~\bibnamefont{Pal}}, \bibnamefont{and}
  \bibinfo{author}{\bibfnamefont{A.~V.} \bibnamefont{Sapre}},
  \bibinfo{journal}{\CPL} \textbf{\bibinfo{volume}{327}}, \bibinfo{pages}{143}
  (\bibinfo{year}{2000}).

\bibitem[{\citenamefont{Ruoff et~al.}(1993)\citenamefont{Ruoff, Malhotra, and
  Huestis}}]{cluster9}
\bibinfo{author}{\bibfnamefont{R.~S.} \bibnamefont{Ruoff}},
  \bibinfo{author}{\bibfnamefont{R.}~\bibnamefont{Malhotra}}, \bibnamefont{and}
  \bibinfo{author}{\bibfnamefont{D.~L.} \bibnamefont{Huestis}},
  \bibinfo{journal}{Nature} \textbf{\bibinfo{volume}{362}},
  \bibinfo{pages}{140} (\bibinfo{year}{1993}).

\bibitem[{\citenamefont{Schweiger and Jeschke}(2001)}]{schweiger2001}
\bibinfo{author}{\bibfnamefont{A.}~\bibnamefont{Schweiger}} \bibnamefont{and}
  \bibinfo{author}{\bibfnamefont{G.}~\bibnamefont{Jeschke}},
  \emph{\bibinfo{title}{Principles of Pulse Electron Paramagnetic Resonance}}
  (\bibinfo{publisher}{Oxford University Press}, \bibinfo{address}{Oxford, UK ;
  New York}, \bibinfo{year}{2001}).

\bibitem[{\citenamefont{Gallagher et~al.}(1996)\citenamefont{Gallagher,
  Armstrong, Lay, and Reed}}]{gallagher96}
\bibinfo{author}{\bibfnamefont{S.~H.} \bibnamefont{Gallagher}},
  \bibinfo{author}{\bibfnamefont{R.~S.} \bibnamefont{Armstrong}},
  \bibinfo{author}{\bibfnamefont{P.~A.} \bibnamefont{Lay}}, \bibnamefont{and}
  \bibinfo{author}{\bibfnamefont{C.~A.} \bibnamefont{Reed}},
  \bibinfo{journal}{\CPL} \textbf{\bibinfo{volume}{248}}, \bibinfo{pages}{353}
  (\bibinfo{year}{1996}).

\bibitem[{\citenamefont{Gallagher et~al.}(1997)\citenamefont{Gallagher,
  Armstrong, Clucas, Lay, and Reed}}]{gallagher97}
\bibinfo{author}{\bibfnamefont{S.~H.} \bibnamefont{Gallagher}},
  \bibinfo{author}{\bibfnamefont{R.~S.} \bibnamefont{Armstrong}},
  \bibinfo{author}{\bibfnamefont{W.~A.} \bibnamefont{Clucas}},
  \bibinfo{author}{\bibfnamefont{P.~A.} \bibnamefont{Lay}}, \bibnamefont{and}
  \bibinfo{author}{\bibfnamefont{C.~A.} \bibnamefont{Reed}},
  \bibinfo{journal}{\JPC~A} \textbf{\bibinfo{volume}{101}},
  \bibinfo{pages}{2960} (\bibinfo{year}{1997}).

\bibitem[{\citenamefont{Hwang and Freed}(1975)}]{hwang75}
\bibinfo{author}{\bibfnamefont{L.-P.} \bibnamefont{Hwang}} \bibnamefont{and}
  \bibinfo{author}{\bibfnamefont{J.~H.} \bibnamefont{Freed}},
  \bibinfo{journal}{\JCP} \textbf{\bibinfo{volume}{63}}, \bibinfo{pages}{4017}
  (\bibinfo{year}{1975}).

\bibitem[{\citenamefont{Freed}(1978)}]{freed78}
\bibinfo{author}{\bibfnamefont{J.~H.} \bibnamefont{Freed}},
  \bibinfo{journal}{\JCP} \textbf{\bibinfo{volume}{68}}, \bibinfo{pages}{4034}
  (\bibinfo{year}{1978}).

\bibitem[{\citenamefont{Polnaszek and Bryant}(1984)}]{polnaszek84}
\bibinfo{author}{\bibfnamefont{C.~F.} \bibnamefont{Polnaszek}}
  \bibnamefont{and} \bibinfo{author}{\bibfnamefont{R.~G.}
  \bibnamefont{Bryant}}, \bibinfo{journal}{\JCP} \textbf{\bibinfo{volume}{81}},
  \bibinfo{pages}{4038} (\bibinfo{year}{1984}).

\bibitem[{\citenamefont{Hinze and Sillescu}(1996)}]{selfdiff1}
\bibinfo{author}{\bibfnamefont{G.}~\bibnamefont{Hinze}} \bibnamefont{and}
  \bibinfo{author}{\bibfnamefont{H.}~\bibnamefont{Sillescu}},
  \bibinfo{journal}{\JCP} \textbf{\bibinfo{volume}{104}}, \bibinfo{pages}{314}
  (\bibinfo{year}{1996}).

\bibitem[{\citenamefont{Assael et~al.}(2001)\citenamefont{Assael, Avelino,
  Dalaouti, Fareleira, and Harris}}]{tolvisc}
\bibinfo{author}{\bibfnamefont{M.~J.} \bibnamefont{Assael}},
  \bibinfo{author}{\bibfnamefont{H.~M.~T.} \bibnamefont{Avelino}},
  \bibinfo{author}{\bibfnamefont{N.~K.} \bibnamefont{Dalaouti}},
  \bibinfo{author}{\bibfnamefont{J.~M. N.~A.} \bibnamefont{Fareleira}},
  \bibnamefont{and} \bibinfo{author}{\bibfnamefont{K.~R.}
  \bibnamefont{Harris}}, \bibinfo{journal}{\IJTP}
  \textbf{\bibinfo{volume}{22}}, \bibinfo{pages}{789} (\bibinfo{year}{2001}).

\bibitem[{\citenamefont{Angell et~al.}(1999)\citenamefont{Angell, Richards, and
  Velikov}}]{cs2glass}
\bibinfo{author}{\bibfnamefont{C.~A.} \bibnamefont{Angell}},
  \bibinfo{author}{\bibfnamefont{B.~E.} \bibnamefont{Richards}},
  \bibnamefont{and} \bibinfo{author}{\bibfnamefont{V.}~\bibnamefont{Velikov}},
  \bibinfo{journal}{\JPCM} \textbf{\bibinfo{volume}{11}}, \bibinfo{pages}{A75}
  (\bibinfo{year}{1999}).

\bibitem[{\citenamefont{Nevzorov and Freed}(2001)}]{nevzorov01}
\bibinfo{author}{\bibfnamefont{A.~A.} \bibnamefont{Nevzorov}} \bibnamefont{and}
  \bibinfo{author}{\bibfnamefont{J.~H.} \bibnamefont{Freed}},
  \bibinfo{journal}{\JCP} \textbf{\bibinfo{volume}{115}}, \bibinfo{pages}{2416}
  (\bibinfo{year}{2001}).

\end{thebibliography}

\end{document}